# Modeling the adoption and use of social media by nonprofit organizations

**Seungah Nah; Gregory D. Saxton**

**Abstract**: This study examines what drives organizational adoption and use of social media through a model built around four key factors – strategy, capacity, governance, and environment. Using Twitter, Facebook, and other data on 100 large US nonprofit organizations, the model is employed to examine the determinants of three key facets of social media utilization: 1) adoption, 2) frequency of use, and 3) dialogue. We find that organizational strategies, capacities, governance features, and external pressures all play a part in these social media adoption and utilization outcomes. Through its integrated, multi-disciplinary theoretical perspective, this study thus helps foster understanding of which types of organizations are able and willing to adopt and juggle multiple social media accounts, to use those accounts to communicate more frequently with their external publics, and to build relationships with those publics through the sending of dialogic messages.

## Introduction

The rapid diffusion of social media applications is ushering in new possibilities for nonprofit organizations to communicate with and engage the public. The ability for any organization – no matter how small – to adopt cutting-edge social media technologies presents substantial opportunities for a more level playing field. It also potentially 'changes the game' with respect to the types of resources and capacities organizations need and the strategies they may adopt in order to successfully capitalize on their social media presence.

Such issues are of critical importance, yet our understanding remains weak. Only a handful of studies (Bortree and Seltzer, 2009; Greenberg and MacAulay, 2009; Waters et al., 2009) have examined any aspect of nonprofits' social media use, predominantly focusing on their efforts at 'dialogic' communication (Kent and Taylor, 1998). The goal of this paper, therefore, is to help boost understanding of what drives organizations to employ social media. To this end, we first propose a comprehensive explanatory model built around four factors – strategy, capacity, governance, and environment – that we posit as key to understanding organizations' adoption and use of social media. Drawing upon Facebook, Twitter, and other data from the 100 largest US nonprofit organizations, we then employ the model to examine the determinants of three key facets of nonprofits' social media use—whether they use it, how frequently they use it, and how often they employ dialogic relationship-building messages.

## Literature Review: Nonprofit Adoption and Use of Social Media

Social media applications have created new ways for organizations to communicate with the public. Twitter and Facebook in particular have garnered attention from nonprofit organizations as innovative communicative tools that both supplement and supplant the traditional Website (Nonprofit Technology Network, 2011). Nonetheless, our understanding of why nonprofits adopt such technologies is sparse not only due to the unique qualities of nonprofit organizations (Lewis, 2005) but also the lack of organizational-level research on social media adoption. There is a substantial intra-organizational communication literature related to *individuals'* adoption, acceptance, and use of new technologies, including the Unified Theory of Acceptance and Use of Technology (Curtis et al., 2010; Venkatesh et al. 2003), the Technology Acceptance Model (e.g., Davis, 1989; Zhou, 2008), Innovation Diffusion Theory (e.g., Rogers, 1995; Vishwanath and Goldhaber, 2003); and process framework (Tang and Ang, 2002). However, such individual-level approaches are better suited to explaining individual preferences for one technology over another—such as why certain employees would prefer Twitter over email—rather than the organizational selection of a given communication technology.

More promising are organizational-level theories of communicative phenomena. First, at a general level, contingency theory (e.g., Cancel et al., 2009) presents a broad framework for explaining organizational variation in public relations strategies (specifically, the level of 'accommodation' with external publics). Though it does not specifically focus on new media or on nonprofit organizations, the approach identifies a large number of factors that could be instructive in the identification of relevant variables.

Second, there are organizational-level Website and information technology adoption studies; those focusing on nonprofit organizations have taken a variety of approaches, including strategic management (Hackler and Saxton, 2007), social and institutional pressures (Zorn, Flanagin, and Shoham, 2011), and other 'organization studies' approaches (Corder, 2001). Such studies are likely useful for explaining social media adoption; however, social media may be *different* from existing technologies in ways that render existing theoretical approaches less relevant. It could be the case, for instance, that budgetary constraints, previously identified as the key factor affecting nonprofits' website and IT adoption (Schneider, 2003; Hackler and Saxton, 2007), may simply no longer be so relevant when it comes to social media.

What little research that does exist on nonprofit social media *utilization,* in turn, regards the heavy reliance on basic informational uses, such as update frequency, as a lost opportunity for deeper engagement with supporters (Bortree and Seltzer, 2009; Greenberg and MacAulay, 2009; Waters et al., 2009). The focus of these studies has been on understanding organizations' 'dialogic' employment of social media (Kent and Taylor, 1998); beyond Waters et al.'s (2009) bivariate analyses of Facebook utilization across industry types, the existing literature does not attempt to explain what drives organizations to adopt social media technologies in the first place, nor what affects the frequency with which nonprofits utilize their social media accounts.

In short, with a literature concentrating on organizations' Website and IT adoption and their dialogic uses of Websites and Facebook, we do not have a good understanding of what compels nonprofits to adopt social media technologies nor what drives them to different communication strategies. Given the dearth of appropriate theory, we weave together elements of the above organizational-level literature into a comprehensive explanatory model, and then use the model to explore the determinants of the three core facets of social media utilization suggested by the literature review: 1) adoption, 2) frequency of use, and 3) dialogic engagement. The following section formally lays out this explanatory model.

## Model and Hypotheses

To help explain nonprofit organizational adoption and use of social media, we propose a theoretical framework that attempts to integrate the factors shown to be key in the existing organizational-level adoption and use literature while addressing key factors from the nonprofit literature that have been understudied in the field of communication. Specifically, we posit that the organizational strategies of the strategic management approach (Hackler and Saxton, 2007), the organizational capacities of the resource mobilization (McCarthy and Zald, 1977) and organization studies (Corder, 2001) frameworks, the governance mechanisms of the management literature (Hambrick and Mason, 1984), and the external pressures of the institutionalism (Zorn et al., 2011) and resource dependence (Pfeffer and Salancik, 1978) schools are all important in determining social media adoption and utilization outcomes.

This framework thus weaves together four dimensions that are critical to understanding an organization's instrumental use of social media. First, organizations choose specific communication strategies in order to achieve their socially driven mission. Second, the ability to successfully reach strategic aims is determined by internal organizational resources and capacities. When nonprofit organizations' preexisting capacities are coupled with the notion of organizational strategy, a set of tools is in place for understanding whether and how nonprofit organizations use social media. However, a third dimension, the organization's governance structures, is essential for ensuring that resources are effectively employed and strategies properly implemented. Finally, the environment in which the organization operates helps drive both the selection and ultimate success of specific communication strategies.

***Strategy***

In nonprofit organizations the ultimate strategic goal is fulfillment of a *social mission*—the creation of public value (e.g., Lewis, 2005). The strategy an organization employs to fulfill this mission has implications for its adoption and use of new media (Hackler and Saxton, 2007). We examine three different strategic approaches to mission fulfillment: fundraising, lobbying, and market-based. Some organizations, such as the Salvation Army, attempt to fulfill their mission via a fundraising focus. Other organizations, such as Greenpeace, focus on lobbying and advocacy to achieve their goals. And others still, such as the YMCA, employ a 'market-based,' fee-for-service strategy for effecting social change. Our first three hypotheses tap organizational variation in these three strategies.

First, a focus on donors, as indicated by fundraising expenses, can be a defining strategic decision (Graddy and Morgan, 2006). Charities following a donor-focused strategy traditionally use mail and telephone solicitations, professional fundraising firms, and special events in order to raise funds. Social media have also recently become a popular fundraising vehicle (Nonprofit Technology Network, 2011). We argue that organizations more focused on acquiring funds through external sources are more likely to adopt and utilize technologies, such as Facebook and Twitter, that enable them to reach and interact with a broader set of potential donors. Our first hypothesis is thus:

*Hypothesis 1: Fundraising expenses will be positively related to social media adoption and use.*

Another way nonprofits seek to fulfill their social mission is through lobbying. Research suggests that, through lobbying and advocacy efforts, nonprofits have enormous potential to improve the lives of their constituents by contributing to democratic governance, influencing public policy, and empowering their constituents to represent themselves effectively (Guo and Musso, 2007; Suárez and Hwang, 2008). Advocacy can hence be seen as not just another service, but as a critical component of a nonprofit's responsibility both to its constituents and to the broader civil society. Organizations following a lobbying strategy may have different communicative needs; we expect politically active nonprofits to be more motivated to use social media, given their interest in mobilizing—often rapidly—a broad external public to take action. To a large extent, the emphasis on a particular strategy is embodied in the amount of resources allocated toward that strategy; accordingly, our second hypothesis is,

*Hypothesis 2: Lobbying expenses will be positively related to social media adoption and use.*

A third approach to effecting social change is to concentrate on market-based program-delivery. Instead of generating revenues through grants or donations, organizations that concentrate on programs generate revenues through market-like fee-for-service transactions, and are thus what Hansmann (1980) calls 'commercial nonprofits.' With a strategy that centers on market-like transactions with clients, we hypothesize such organizations have a greater incentive to reach out to both current and existing customers through social media:

*Hypothesis 3: Program Service revenues will be positively related to social media adoption and use.*

***Capacity***

The capacity and resources an organization can mobilize (McCarthy and Zald, 1977) in pursuit of strategically driven initiatives has implications for the adoption and use of social media. We propose three capacity-related factors. Our first proxy for capacity is organizational size as reflected in total financial assets. To start, size affects the acquisition of new technology (Corder, 2001; Zorn et al., 2011). Moreover, as an organization grows, it becomes more visible and therefore attracts greater attention and scrutiny by external constituencies such as the state, the media, and the general public (Luoma and Goodstein, 1999). This in turn may lead organizations to a larger social media presence to address these stakeholders' concerns. Size is also consistently a critical factor in determining both access to technology and the general 'IT capacity' of nonprofit organizations (Hackler and Saxton, 2007; Schneider, 2003). Finally, the use of social media is not cost-free—organizations with successful social media efforts must devote resources in terms of time and money—and larger organizations are better able to afford the investment. We thus make the following hypothesis:

*Hypothesis 4: Organizational size in assets will be positively related to social media adoption and use.*

We then incorporate two characteristics of the organization's existing Website capabilities and presence. There are several arguments that imply a positive relationship for these factors with social media use. First, in line with diffusion of innovations theory (Rogers, 1995), older websites are indicative of earlier Website adopters; this might indicate an organization that is also more likely to be an earlier adopter of social media.

Second, organizations with a substantial Website presence might feel pressure to use newer and more advanced technologies such as Twitter and Facebook. This could come through competitive forces, as organizations strive to maintain a new media-driven communicative competitive advantage (Porter, 1985); alternatively, the pressure might come from their large online user bases to continue to adopt newly emergent digital communication technologies.

Third, as resource mobilization theory would imply (McCarthy and Zald, 1977), preexisting Web capabilities might constitute resources that organizations can mobilize in pursuit of additional Web-based goals (Kropczynski and Nah, 2011). There is in fact growing evidence that Internet and Website capacities constitute critical organizational capabilities for the successful strategic use of information technology (Hackler and Saxton, 2007). Website reach is also an indirect indicator of 'communication competency,' which in contingency theory (Cancel et al., 2009) is posited as a determinant of an organization's external trust-building efforts.

In line with the above arguments, we hypothesize that organizations with a more established and far-reaching Web presence, as measured by the age of and number of visitors to their Website, should be more likely to adopt and use social media:

*Hypothesis 5: Website age will be positively related to social media adoption and use.*

*Hypothesis 6: Website reach will be positively related to social media adoption and use.*

***Governance***

The upper-echelons perspective (Hambrick and Mason, 1984) attributes major influence to organizational governance, and it has been found to play an important role in nonprofits' adoption of Web technologies (Saxton and Guo, 2011). We thus develop three hypotheses to capture key elements of organizations' governance characteristics.

Our first measure of governance captures a particular 'type' of governance structure, that embodied in membership organizations. Membership-based nonprofits are in important ways different from the average nonprofit organization (Smith, 1993). There are clearly defined organizational boundaries and areas of stakeholder concern (members vs. non-members). Especially relevant is that, unlike non-membership organizations, where the leadership is self-perpetuating, membership organizations have a more bottom-up, representative governance structure, and usually offer opportunities for members to partake in direct elections on strategic and leadership matters. Guo and Musso (2007) suggest that where such formal modes of representation are available, mechanisms of stakeholder communication and participation, such as that achieved by social media, are less urgent. Given this bottom-up structure and less intense need to reach a broad spectrum of external publics, we posit top-down, organizational-driven communication on social media will not be as prevalent:

*Hypothesis 7: Membership-based nonprofit organizations will be less likely to adopt and use social media.*

Our next hypothesis relates to an 'input' into the organization's governance: board size, a commonly employed measure of governance. To start, larger boards are more likely to have a social media 'champion' present, which prior research suggests is strongly connected to IT adoption (Howell and Higgins, 1990). More importantly, larger boards generally have more contact with the public, which facilitates fundraising and other externally driven activities (Olson, 2000). In effect, a larger board indicates greater external ties, which could spur the use of social media to solidify those ties. Consequently, though larger boards are not always better, we hypothesize a positive association between board size and social media utilization:

*Hypothesis 8: Organizations with larger boards will be more likely to adopt and use social media.*

We further propose that better governed organizations, on account of their more participatory and/or effective stakeholder involvement and communication strategies (Guo and Musso, 2007) will be more likely to employ social media. Our proxy for effective governance is financial stewardship, a core board responsibility (Gill, Flynn, and Reissing, 2005: 278). A common indicator of financial stewardship is the well known 'program spending ratio,' an efficiency measure defined as the ratio of program expenses to total expenses (see Parsons, 2003, for an overview). This leads to the following hypothesis:

*Hypothesis 9: More efficient nonprofits will be more likely to adopt and use social media.*

***Environment***

The final determinant in our model focuses on organizations' external resource environment. This factor encompasses the pressures to adopt new technologies generated by external constituents and social and institutional forces (Corder, 2001; Zorn et al., 2011). It also reflects the ideas of resource dependence theory (Pfeffer and Salancik, 1978), wherein an organization's behavior is conditioned by the extent to which resources critical for its survival are controlled by actors in its external environment. This view of stakeholder relations holds that power, and in turn managerial attention, shifts to those stakeholders who control critical resources (e.g., Mitchell, Agle, and Wood, 1997). We capture these ideas via *Donor Dependence* and *Government Dependence,* which measure the proportion of organizational revenues derived from public contributions and government funding, respectively. We hypothesize that, the higher the level of dependence on these external stakeholders, the greater the level of online attention will be afforded them, and by extension, the greater the likelihood the organization will employ social media in its stakeholder relations efforts:

*Hypothesis 10: Donor dependence will be positively related to an organization's social media adoption and use.*

*Hypothesis 11: Government dependence will be positively related to an organization's social media adoption and use.*

## Method

### *Sample*

Given the relative novelty of social media, our sample comprises large charitable organizations, which are more likely than smaller organizations to have a significant social media presence. Specifically, as in prior studies (e.g., Kang and Norton, 2004; Yeon, Choi, and Kiousis, 2007), we examined organizations from the most recent version (2008) of the '*Nonprofit Times* 100' list available at the start of our study period. Published annually in the *NonProfit Times*, the list contains the 100 largest non-educational US nonprofit organizations in terms of revenue. Appendix A contains a complete list of the organizations with associated industry codes.

### *Data Collection and Measurement*

Data were gathered from multiple sources, including, for the independent variables, the National Center for Charitable Statistics, 2008 IRS tax-return forms, the Internet Archive, and Google; and, for the dependent variables, Twitter and Facebook. We discuss data-gathering and measurement of the independent variables before turning to the dependent variables. Appendix B contains descriptive statistics and zero-order correlations for all model variables.

*Independent Variables*

Using publicly available 2008 IRS Form 990 data, we first measured three variables for organizational strategy: *Fundraising Ratio,* the ratio of fundraising expenses to total expenses; *Lobbying Expenses,* the amount spent on grassroots and direct lobbying activities (in $10,000); and *Program Service Revenues,* the amount (in $Million) generated via the delivery of programs.

We then operationalized three measures of organizational capacity. *Size* was measured as net total assets in 2008, as reported in IRS Form 990 data. To adjust for the skewed distribution of the variable, we transform the variable by taking the natural logarithm. *Website Age* was measured using the date of the organization's first appearance in the Internet Archive Wayback Machine (www.archive.org), a source validated by scholarly research (e.g., Murphy, Hashim, and O'Connor, 2008). *Website Reach*, in turn, was measured as the number of 'inlinks' reported on Google. For instance, typing in 'link:www.foodforthepoor.org' in the Google Search engine returned 851 links from external Websites, not including blogs. It is a general measure of the degree of influence of the organization's Website.

Our third explanatory factor is governance. We measured three variables using 2008 IRS Form 990 data. First, *Membership* is a dichotomous variable reflecting whether the organization was membership-based, as indicated by the presence of membership funding. *Board Size* measures the number of people on the board of directors, while *Efficiency* is measured as the proportion of total expenses devoted to programs (program expenses/total expenses).

Lastly, we operationalize the organization's resource environment via measures of reliance on donations and government funding, respectively. Using 2008 IRS 990 data, *Donor Dependence* was measured as the ratio of income from public support to total income; while *Government Dependence* was measured as the ratio of income from government to total income.

*Control Variables*

We also coded control variables for the organizations' age and industry type. Both are noted in Contingency Theory (Cancel et al., 1997) as potential influencers of 'accommodation' strategies, and are standard control variables in nonprofit studies (e.g., Waters et al., 2009). We measured *Age* as the number of years since the organization was officially recognized as a tax-exempt organization. This information is available on the National Center for Charitable Statistics Website. Second, using IRS-provided NTEE codes from the National Center for Charitable Statistics (nccsdataweb.urban.org), we created dummy variables to account for three specific industries or fields of interest: *Arts* (n = 14), *Health* (n = 21), and *Human Services* (n = 14).

*Dependent Variables*

Social media utilization was measured with three dimensions—presence, volume, and dialogue—which collectively tap whether and to what extent nonprofit organizations use social media, specifically Facebook and Twitter, to engage the public. First, through a comprehensive search in November of 2009 on Facebook, Twitter, Google, and organizational websites, we found that 73 of the 100 organizations had Twitter accounts and 65 Facebook accounts. Only 53 organizations had adopted both Twitter and Facebook, while fully 15 had no social media presence. Selecting a period long enough (one month) to ensure an adequate number of status updates, we then gathered Twitter and Facebook data for each organization. For Twitter, all organizational tweets published from November 8$^{th}$ to December 7$^{th}$, 2009 were downloaded into an *SQLite* relational database via the Twitter application programming interface (API), using Python code written for this research (available upon request). The final database contained 4,655 tweets, which were double-checked against the Twitter stream for 10 of the organizations and found to be complete in all cases. A subset of these tweets—the 2,437 sent over the first two weeks of the study—were then hand coded according to whether the primary purpose of the tweet was informational, promotional, or dialogic in intent. 'Dialogic' tweets were those that conformed to Kent and Taylor's (1998) notion of the 'dialogic loop' inasmuch as they signal an intent to query, converse, and build relationships with an external public. Inter-coder agreement for two investigators for 200 of the 2,437 tweets was 94.0%, with a Cohen's kappa score of .91.

Original Python code was similarly used to download all 1,036 status updates from the organizations' Facebook pages during the period December 5, 2009 to January 4, 2010. To ensure Python was downloading the updates accurately, we conducted a trial download and randomly compared 100 downloaded status updates with their counterpart updates on the Facebook website. Status updates were downloaded correctly in all cases. As with the tweets, each of the 1,036 statuses was then hand-coded to determine whether the primary purpose of the message involved dialogic relationship-building. Inter-coder agreement for 157 of the 1,036 statuses was 93.2% with a Cohen's Kappa score of .89.

Using these data we created nine dependent variables. First, each organization's social media presence was coded with three variables: *Facebook Presence* and *Twitter Presence*, binary variables (0,1) indicating organizational adoption of Facebook and Twitter, respectively; and *Social Media Presence*, with values of '0' for organizations on neither Twitter nor Facebook, '1' for organizations on either Twitter or Facebook alone, and '2' for those organizations on both Facebook and Twitter.

We then created three variables that tap the frequency with which organizations used their social media applications over the month-long study period: *Frequency of Facebook Updates,* a count of the organization's Facebook status updates; *Frequency of Twitter Updates,* a count of the number of Twitter updates (tweets) posted by the organization; and *Frequency of Social Media Updates,* the combined total of each organization's Twitter and Facebook updates.

Lastly, we created three variables to tap the frequency with which an organization engages in external relationship-building through the sending of dialogic messages. *Dialogic Facebook Messages* measures the number of each organization's Facebook statuses that were dialogic in nature, *Dialogic Twitter Messages* measures the number of dialogic tweets, and *Dialogic Social Media Messages* constitutes a combined count of the number of dialogic messages sent on Twitter and Facebook.

### *Analytical Techniques*

Because our first two dependent variables (*Facebook Presence* and *Twitter Presence*) are binary categorical variables, the use of ordinary least squares (OLS) would result in biased, inefficient, and inconsistent parameter estimates (Long, 1997). We thus employ logit regressions to estimate these two models. Our third dependent variable, *Social Media Presence,* is measured on an ordinal categorical scale; accordingly, we employ a maximum likelihood technique, generalized ordered logit (Williams, 2006), to test the relationship between our explanatory variables and social media presence. The six remaining dependent variables are ratio-level count variables. As with most count data, these variables have a non-normal distribution that includes a high number of low-frequency occurrences. The OLS method would as a result produce inaccurate parameter estimates (Long, 1997). To deal with this problem, researchers typically employ various nonlinear models based on the Poisson and negative binomial distributions. In our case, the dispersion for the six count variables is greater than would be expected for a traditional Poisson distribution (the variances are much larger than the means), and we thus estimate the models using a negative binomial estimation technique.

## Results

As Table 1 indicates, each regression model obtains significant Chi-squared values, and pseudo-$R^2$ values range between .18 and .43. We summarize findings by hypothesis in order. First, hypothesis 1, which had predicted a positive relationship between the *Fundraising Ratio* and social media utilization*,* received no support. The variable obtained significance in three models (*Frequency of Twitter Updates, Frequency of Social Media Updates,* and *Dialogic Facebook Messages*), but in the opposite direction from that expected.

[Table 1 about here]

Hypothesis 2 posited a positive relationship between *Lobbying Expenditures* and social media utilization. As expected, the variable obtains a positive and significant coefficient in Model 1, indicating a positive association between the amount an organization engages in lobbying and its adoption of Facebook. However, in the four other models the variable obtains significance (*Social Media Presence, Frequency of Facebook Updates*, *Dialogic Twitter Messages,* and *Dialogic Social Media Messages*), the coefficient is negative. This suggests that, overall, lobbying organizations are less likely to adopt social media, less likely to send social media updates, and less likely to send dialogic messages to engage their external publics.

The coefficient on the measure of *Program Service Revenues* was significant and positive in four of the nine models, those for *Twitter Presence, Social Media Presence, Frequency of Facebook Updates,* and *Frequency of Social Media Updates* (it was insignificant in the others). As expected by hypothesis 3, 'commercial' nonprofits (those with a stronger fee-for-service model) are more likely to have a social media presence and to engage clients through more frequent social media messages.

Hypotheses 4 through 6 posited relationships between social media utilization and organizational capacity. *Size* yielded a significant, yet negative relationship only with the *Twitter Presence* variable. Hypothesis 4 thus receives no support. Hypothesis 5, positing a positive relationship between *Website Age* and social media utilization, is similarly unsupported, given that it obtained a significant (and negative) association in only one model (*Facebook Presence*). However, *Website Reach* was one of the most robust variables in the model. It yielded significantly strong and positive relationships with all but one of the dependent variables (*Facebook Presence*). Hypothesis 6 therefore receives substantial support.

Our first governance variable, *Membership,* was found to obtain a significant and positive association with the *Facebook Presence* variable (Model 1), and a negative association with two of the three volume variables (*Frequency of Twitter Updates* and *Frequency of Social Media Updates*) as well as all three 'relationship-building' variables (Models 7-9). Therefore, hypothesis 7, which posited a negative relationship between membership-based organizations and social media adoption and use, is partially supported. Hypotheses 8 and 9, meanwhile, posited positive relationships with social media utilization for *Board Size* and *Efficiency,* respectively. Hypothesis 8 receives only marginal support: though the *Board Size* coefficient was significant and positive in Model 2 (*Twitter Presence*), it was negative in three of the models (Models 4, 8, and 9) and insignificant in the others. There is support for Hypothesis 9, however. The coefficient on *Efficiency* obtained significance with five of the nine dependent variables (*Facebook Presence*, the three frequency variables, and *Dialogic Facebook Messages*).

Hypotheses 10 and 11 concerned organizations' reliance on, respectively, donations and government funding. *Donor Dependence* obtained a significant positive association with all three adoption variables (Models 1-3), with *Frequency of Facebook Updates* (Model 4), and with *Dialogic Facebook Messages* (Model 7). However, it was significantly negative with *Frequency of Twitter Updates* (Model 5), *Dialogic Twitter Messages* (Model 8), and *Dialogic Social Media Messages* (Model 9). *Government Dependence*, meanwhile, obtained a statistically significant (negative) relationship only in the regression on the number of *Dialogic Facebook Messages* (Model 7). In effect, hypothesis 10 receives partial support and hypothesis 11 no support.

## Discussion and Conclusions

This study proposed and examined an integrated approach to understanding how nonprofit organizations adopt and use social media. Drawing upon data from the 100 largest US nonprofits, we find that organizational strategies, capacities, governance features, and external pressures all play a part in social media adoption and utilization outcomes.

Our first set of variables tapped variation in organizational strategies. We found that *Lobbying expenditures* had an inconsistent effect, but was generally associated with less frequent social media updating and the sending of dialogic messages. Second, *Program Service Revenues* were generally positively associated with social media adoption and the volume of updates. This finding suggests that nonprofit organizations that focus strategically on obtaining revenues from market-based program-delivery rather than grants or donations tend to rely more on social media to facilitate communications with their clients. Third, contrary to expectations, we found that fundraising was negatively related to how frequently the organizations actually used social media in terms of both message volume and engaging in dialogue. It might be the case that, as an organization becomes accustomed to using social media to communicate with key stakeholders, social media use may 'crowd out' more costly 'offline' fundraising activities. It is in this sense that traditional, paid fundraising activities and the sending of social media messages may actually be 'substitutes.' These ideas are worth exploring in future research.

Similarly, organizational utilization of social media appears to depend on pre-existing resources and capacities, especially those related to Web capabilities (e.g., Hackler and Saxton, 2007). How long an organization had maintained a website was shown to be largely insignificant. However, pre-existing Website reach proved to be a powerful predictor of social media utilization. Ostensibly, the capacities an organization builds up in order to develop a more influential Website pays dividends when it comes to the ability to adopt and effectively utilize newly emergent technologies such as Facebook and Twitter.

Surprisingly, size in assets was not positively related to any of the dependent variables, implying that size does not represent a barrier to the employment of social media, similar to what Nah (2010) and Yeon, Choi, and Kiousis (2007) found in their studies of nonprofit Website use. Yet this is distinct from what prior research has found regarding other forms of technology, such as access to computers and information technology (Hackler and Saxton, 2007; Schneider, 2003). However, especially given the nature of our 'large organization' sample, we need to be cautious in interpreting findings related to the assets–social media nexus. Future studies should continue to explore the relationship.

Our measures of organizational governance were also shown to play vital roles in predicting social media utilization. The organization's governance characteristics are crucial in ensuring that resources are effectively employed and strategies properly implemented with respect to the adoption and use of information technology (Hackler and Saxton, 2007). In our study, membership structure (Smith, 1993), board size (Olson, 2000), and organizational efficiency (Gill et al., 2005) were found to be key determinants of social media adoption and use. These findings are consistent with previous scholarship on organizational governance and leadership as a key factor in the adoption of new communication technologies (e.g., Saxton and Guo, 2011). Our research has highlighted several effects that are ripe for future research.

Specifically, *Membership* was found to be strongly related to decreased volume of social media updating and dialogic outreach. Not only do member organizations have a distinct, more participatory governance structure, they have a more clearly defined stakeholder focus. More research should be done to explicitly connect participatory organizational mechanisms and stakeholder targeting to social media outcomes. Next, *Board Size* had an inconsistent effect, being associated with increased likelihood of Twitter adoption but lower volumes of social media updating and the sending of dialogic messages. *Efficiency,* meanwhile, though not always significant, had a consistently positively association with social media adoption and use.

Beyond such internal variables, we also found external factors played significant roles in predicting nonprofits' adoption and use of social media. This finding is consistent with previous studies arguing that pressures from external stakeholders (e.g., Corder, 2001; Mitchell et al., 1997; Pfeffer and Salancik, 1978; Zorn et al., 2011) can push nonprofit organizations to implement new technologies. Specifically, we found reliance on public donations has a considerable effect in influencing how organizations employ social media. Reliance on government funding, in contrast, was not shown to be important; the results are hence too weak to determine whether government funding 'pushes' organizations one way or the other. Overall, what do these results mean for the resource dependence perspective? Primarily, they show that some resources are more important than others. What we found here was that, controlling for other factors, including the organization's assets, the degree of dependence on donors matters much more than the degree of dependence on funding by the government. In retrospect, this might make sense: organizations' funding from governmental sources may rely more on an organizations' grant-writing and performance-measurement capacities, whereas obtaining funding from donors requires a more extensive outreach, stakeholder engagement, and public relations focus. It is thus perhaps not surprising that dependence on donors appears to push organizations to use social media more heavily. Accordingly, we strongly suggest future research incorporates this variable into their explanations.

There were differences between Facebook and Twitter in terms of the effects of the various predictors of the adoption and use of these two social media. Specifically, the sign of the coefficients was different for 6 of the 11 variables in the individual-media 'presence' tests (Models 1 and 2), for 3 of the variables in the 'volume' tests (Models 4 and 5), and for 4 of the variables in the 'dialogue' tests (Models 7 and 8). These findings suggest that scholars should be cautious in making claims about the determinants of 'social media' in general. Facebook and Twitter are in some ways different tools that can be used for different purposes, and may as a result require different configurations of organizational resources, governance characteristics, and contextual and environmental factors in order to implement and maintain them.

These findings carry important broader theoretical implications beyond those of the individual variables and factors noted above. Our study examines two different types of social media, Facebook and Twitter, with not only the presence and frequency of updating messages but also frequency of sending specific kinds of messages—those intended to foster dialogic communication. The insights are thus applicable to the broad literatures on technology adoption, social media use, dialogue, and organization-public relations. Moreover, our study adds insights to a relatively unexplored area through an integrated, multi-disciplinary theoretical perspective that aims to help uncover what drives organizational adoption and use of social media. Overall, our study helps foster understanding of which types of organizations are able and willing to adopt and juggle multiple social media accounts, to use those accounts to communicate more frequently with their external publics, and to build relationships with those publics through the sending of dialogic messages.

There are also practical implications for both social media ‘user’ organizations and for grantmaking and community-based ‘capacity-building’ organizations. For ‘using’ organizations, our study contributes to an understanding of the factors that influence the successful implementation and employment of social media. For grantmaking and capacity-building organizations, the findings shed light on where nonprofits in their community need help.

There are also several limitations and suggestions for future research. First, the fact that the variance explained by several previously established variables (such as size) appears to be weaker in these tests than in studies of prior technologies suggests there might be something different about social media that has ‘freed’ nonprofits from some of the capacity and environmental constraints that have hampered them in the past. More broadly, the results suggest that future versions of the model should consider additional variables, especially the organization’s ‘human resources’ as well as such ‘public relations’ variables as the organization’s public relations model and its susceptibility to crisis. Second, although the results from these 100 large organizations provide useful insights, and facilitate comparisons to studies conducted using prior versions of the *NPTimes* 100 lists (e.g., Kang and Norton, 2004; Yeon, Choi, and Kiousis, 2007), they are not necessarily generalizable to midsize and small nonprofit organizations. This calls for studies of randomized samples of a broader range of organizations. Future studies should also consider employing other methodological approaches, including case studies, surveys, ethnography, and in-depth interviews, to provide more in-depth explanations of how and why nonprofit organizations adopt and use social media.

Despite the limitations, this study adds important insights to the relationship between nonprofit organizations and their utilization of social media with its focus on a set of core internal and external organizational factors. This study verifies that strategy, capacity, governance, and environment play a key role in mobilizing newly emerging social media as alternative and additional communication tools that help nonprofit organizations to strategically maintain and maximize resources. As a consequence, this study provides a theoretical framework that future studies can expand, develop, challenge, and test using different samples and different methodological approaches. In so doing, we can gain a more nuanced understanding of how organizations can better adapt and mobilize their existing governance structures, internal capacities and external environment in support of communication strategies that can achieve success in the rapidly changing social media environment.

Table 1. Analyses of Social Media Adoption and Use by *NPTimes 100* Nonprofit Organizations

| | **Presence of Social Media** | | | **Frequency of Social Media Use** | | | **Frequency of Dialogic Messages** | | |
|---|---|---|---|---|---|---|---|---|---|
| | (Model 1) Facebook Presence (0,1) | (Model 2) Twitter Presence (0,1) | (Model 3) Social Media Presence (0,1,2) | (Model 4) Frequency of Facebook Updates | (Model 5) Frequency of Twitter Updates | (Model 6) Frequency of Social Media Updates | (Model 7) Dialogic Facebook Messages | (Model 8) Dialogic Twitter Messages | (Model 9) Dialogic Social Media Messages |
| *Strategy* | | | | | | | | | |
| Fundraising Ratio | 5.63 (6.46) | -3.31 (8.88) | 1.96 (6.57) | -3.62 (2.36) | -4.81$^{+}$ (2.52) | -3.45$^{+}$ (2.03) | -4.68$^{+}$ (2.81) | -4.52 (3.22) | -3.92 (3.08) |
| Lobbying Expenses | .004$^{+}$ (.002) | -.001 (.001) | -.03$^{*}$ (.01) | -0.01$^{**}$(.001) | -.0003 (.001) | -.0006 (.0004) | -.0002 (.002) | -.002$^{**}$(.007) | -.002$^{+}$(.001) |
| Program Serv. Rev. | .0002 (.002) | 0.01$^{*}$(.004) | .005$^{**}$(.0002) | .002$^{**}$(.001) | -.0001 (.0001) | .002$^{*}$(.0006) | .0002 (.0004) | -.0002 (.0004) | -.0001 (.0004) |
| *Capacity* | | | | | | | | | |
| Size in Assets | -0.20 (0.19) | -0.46$^{*}$(0.19) | -0.23 (0.19) | -0.06 (0.07) | -0.01 (0.09) | -0.08 (0.07) | -0.08 (0.03) | -0.14 (0.14) | -0.03 (0.11) |
| Website Age | -0.34$^{+}$ (0.18) | 0.06( 0.14) | -0.20(0.16) | -0.03(0.08) | -0.01(0.05) | 0.00(0.04) | -0.09(0.11) | 0.00(0.07) | -0.01(0.07) |
| Website Reach | 0.30 (0.77) | 1.97$^{*}$(0.91) | 1.43$^{+}$(0.77) | 1.47$^{**}$(0.30) | 1.27$^{**}$(0.36) | 1.63$^{**}$(0.23) | 1.48$^{**}$(0.42) | 1.18$^{**}$(0.43) | 1.29$^{**}$(0.34) |
| *Governance* | | | | | | | | | |
| Membership org | 1.71$^{+}$ (1.06) | -0.67(0.88) | 0.50(0.61) | -0.52(0.39) | -1.12$^{*}$(0.48) | -1.07$^{**}$(0.34) | -1.62$^{**}$(0.55) | -1.12$^{*}$(0.54) | -1.15$^{**}$(0.45) |
| Board Size | -0.00 (0.01) | 0.02$^{+}$(0.01) | 0.00(0.01) | -0.01$^{*}$(0.01) | -0.01(0.01) | -0.01(0.00) | -0.01(0.01) | -0.01$^{+}$(0.01) | -0.01$^{*}$(0.01) |
| Efficiency | 7.16$^{*}$ (3.17) | 0.13(0.14) | 0.80(3.75) | 0.07$^{+}$(0.04) | 0.08$^{*}$(0.04) | 0.06$^{*}$(0.03) | 0.10$^{*}$(0.04) | 0.06(0.05) | 0.07(0.04) |
| *Environment* | | | | | | | | | |
| Donor Dependence | 0.67$^{+}$ (0.38) | 2.68$^{*}$(1.34) | 1.68$^{+}$(0.93) | 0.31$^{**}$(0.12) | -0.30$^{*}$(0.13) | 0.11(0.16) | 0.34$^{+}$(0.19) | -0.92$^{**}$(0.31) | -0.31$^{*}$(0.16) |
| Gov't. Dependence | 0.76 (1.19) | -0.61(1.36) | -0.05(1.25) | -0.15(0.55) | 0.19(0.59) | 0.08(0.54) | -1.67$^{+}$(0.91) | 0.00(0.73) | -0.21(0.74) |
| Industry/Age controls? | YES | YES | YES | YES | YES | YES | YES | YES | YES |
| *N* | 97 | 97 | 97 | 63 | 71 | 82 | 63 | 71 | 82 |
| Pseudo $R^2$ | 0.26 | 0.22 | 0.18 | 0.34 | 0.37 | 0.43 | 0.29 | 0.37 | 0.32 |
| Log likelihood | -51.49 | -43.04 | -77.80 | -225.73 | -354.39 | -404.01 | -120.95 | -205.44 | -250.31 |
| Model Significance ($\chi^2$) | 25.78$^{*}$ | 39.43$^{**}$ | 44.46$^{**}$ | 25.76$^{*}$ | 32.39$^{**}$ | 45.83$^{**}$ | 21.79$^{+}$ | 32.84$^{**}$ | 31.92$^{**}$ |

*Notes:* Logistic/Negative Binomial regression coefficients shown for all models, with robust standard errors in parentheses. $^{+}$ $p < 0.10$, $^{*}$ $p < 0.05$, $^{**}$ $p < 0.01$
McFadden $R^2$ shown for logit regressions, ML (Cox-Snell) $R^2$ shown for negative binomial regressions.
*Industry/Age controls?* - Indicates that age and industry control variables are included in the regression.
*Models 1-2:* Logit regression.

*Model 3:* A Brant Test showed the model violated the parallel regression assumption, rendering typical ordered logit regression unsuitable; we thus employed a generalized ordered logit model, which does not require the parallel lines assumption, using Williams' (2006) gologit2 procedure.

*Models 4-9:* Negative binomial regression. Likelihood ratio (LR) tests show the negative binomial model provides a significantly better fit than a Poisson model. The α values for the LR tests for each of our negative binomial regressions are significantly different from 0, indicating the inappropriateness of a Poisson model, which assumes that $\alpha = 0$.

Appendix A

*List of Sample Organizations, with NTEE Codes and Social Media Data*

| *NP Times* 100 Rank | Organization | Field (NTEE code) | On Twitter? | On Facebook? | # Tweets | # FB Updates | # Dialogic (Twitter) | # Dialogic (FB) |
|---|---|---|---|---|---|---|---|---|
| 1 | YMCA of the USA | P | no | yes | . | 0 | . | 0 |
| 2 | Catholic Charities USA | P | yes | yes | 6 | 4 | 1 | 0 |
| 3 | The Salvation Army | P | yes | yes | 139 | 27 | 2 | 3 |
| 4 | Goodwill Industries International | J | yes | yes | 103 | 29 | 14 | 7 |
| 5 | American Red Cross | P | yes | yes | 22 | 24 | 3 | 4 |
| 6 | Boys & Girls Clubs of America | O | no | yes | . | 12 | . | 1 |
| 7 | Habitat for Humanity International | L | yes | no | 67 | . | 4 | . |
| 8 | National Easter Seal Society | E | yes | no | 27 | . | 2 | . |
| 9 | American Cancer Society | G | yes | yes | 208 | 5 | 5 | 0 |
| 10 | Food For The Poor | Q | yes | yes | 44 | 27 | 4 | 5 |
| 11 | The Nature Conservancy | Q | yes | yes | 174 | 45 | 10 | 15 |
| 12 | Planned Parenthood Federation of America | E | yes | no | 92 | . | 3 | . |
| 13 | Boy Scouts of America | O | yes | yes | 22 | 32 | 1 | 6 |
| 14 | World Vision | Q | yes | yes | 289 | 77 | 1 | 19 |
| 15 | Feed the Children | Q | yes | yes | 13 | 9 | 0 | 2 |
| 16 | Shriners Hospitals for Children | E | yes | yes | 14 | 15 | 0 | 3 |
| 17 | AmeriCares Foundation | Q | yes | no | 167 | . | 27 | . |
| 18 | Volunteers of America | P | yes | yes | 46 | 28 | 2 | 2 |
| 19 | Girl Scouts of the USA | O | yes | yes | 15 | 1 | 1 | 0 |
| 20 | Dana Farber Cancer Institute | G | yes | yes | 11 | 7 | 0 | 1 |
| 21 | Gifts In Kind International | T | yes | yes | 20 | 1 | 1 | 0 |
| 22 | ALSAC/St Jude's Children's Research Hospital | E | yes | no | 38 | . | 4 | . |
| 23 | City of Hope and affiliates | E | yes | no | 9 | . | 0 | . |
| 24 | Feeding America | K | yes | yes | 36 | 19 | 3 | 2 |
| 25 | American Heart Association | G | yes | yes | 27 | 0 | 1 | 0 |
| 26 | CARE | Q | yes | no | 260 | . | 38 | . |
| 27 | Public Broadcasting Service | A | yes | no | 202 | . | 39 | . |
| 28 | Children's Hospital Los Angeles | E | yes | no | 90 | . | 16 | . |
| 29 | Catholic Relief Services | M | yes | yes | 34 | 19 | 2 | 4 |
| 30 | Smithsonian Institution | A | yes | yes | 120 | 24 | 17 | 0 |
| 31 | Campus Crusade for Christ International | O | yes | no | 5 | . | 0 | . |
| 32 | United Cerebral Palsy Associations | G | yes | no | 48 | . | 5 | . |
| 33 | Metropolitan Museum of Art | A | yes | yes | 56 | 13 | 4 | 2 |
| 34 | Eisenhower Medical Center and affiliates | E | no | no | . | . | . | . |
| 35 | Academy for Educational Development | Q | no | no | . | . | . | . |
| 36 | Map International | Q | yes | yes | 25 | 4 | 1 | 2 |
| 37 | Scripps Research Institute | H | no | no | . | . | . | . |
| 38 | New York Presbyterian Fund | E | no | no | . | . | . | . |
| 39 | Save the Children Federation | Q | yes | no | 44 | . | 3 | . |
| 40 | US Fund for UNICEF | Q | no | no | . | . | . | . |
| 41 | Brother's Brother Foundation | Q | no | yes | . | 0 | . | 0 |
| 42 | Pew Charitable Trusts | T | no | no | . | . | . | . |
| 43 | JUF/Jewish Federation of Metropolitan Chicago | T | yes | no | 43 | . | 5 | . |
| 44 | Compassion International | Q | yes | yes | 69 | 49 | 8 | 7 |
| 45 | Museum Of Modern Art | A | yes | yes | 73 | 28 | 24 | 4 |
| 46 | Hadassah | Q | yes | yes | 2 | 0 | 0 | 0 |
| 47 | New York Public Library | B | yes | yes | 136 | 24 | 11 | 1 |
| 48 | Samaritan's Purse | Q | yes | yes | 19 | 0 | 2 | 0 |
| 49 | Susan G Komen For the Cure | G | yes | yes | 57 | 11 | 7 | 4 |
| 50 | Big Brothers/Big Sisters of America | O | yes | yes | 4 | 8 | 0 | 0 |

| *NP Times* 100 Rank | Organization | Field (NTEE code) | On Twitter? | On Facebook? | # Tweets | # FB Updates | # Dialogic (Twitter) | # Dialogic (FB) |
|---|---|---|---|---|---|---|---|---|
| 51 | UJA Federation of Jewish Philanthropies of NY | T | yes | no | 25 | . | 1 | . |
| 52 | American Jewish Joint Distribution Committee | Q | no | no | . | . | . | . |
| 53 | Metropolitan Opera Association | A | yes | no | 41 | . | 16 | . |
| 54 | International Rescue Committee | Q | yes | yes | 164 | 27 | 42 | 1 |
| 55 | Art Institute of Chicago | B | yes | yes | 73 | 16 | 6 | 1 |
| 56 | Institute of International Education | Q | no | yes | . | 2 | . | 1 |
| 57 | Leukemia & Lymphoma Society | G | yes | yes | 40 | 9 | 1 | 1 |
| 58 | Christian Broadcasting Network | X | no | no | . | . | . | . |
| 59 | March of Dimes Foundation | G | yes | yes | 146 | 22 | 29 | 9 |
| 60 | Special Olympics | N | yes | yes | 43 | 13 | 1 | 1 |
| 61 | Cystic Fibrosis Foundation | G | yes | yes | 80 | 13 | 22 | 0 |
| 62 | Direct Relief International | Q | yes | yes | 47 | 11 | 5 | 1 |
| 63 | National Multiple Sclerosis Society | G | no | yes | . | 15 | . | 1 |
| 64 | Alzheimer's Association | G | no | yes | . | 7 | . | 0 |
| 65 | National Gallery of Art | A | no | yes | . | 12 | . | 0 |
| 66 | Wildlife Conservation Society | D | no | no | . | . | . | . |
| 67 | Cross International Aid/Catholic Outreach | Q | no | yes | . | 8 | . | 2 |
| 68 | Operation Blessing International Relief | Q | yes | no | 70 | . | 6 | . |
| 69 | American Diabetes Association | G | yes | yes | 51 | 7 | 16 | 1 |
| 70 | World Wildlife Fund | D | yes | yes | 12 | 13 | 0 | 2 |
| 71 | Boys Town | P | no | no | . | . | . | . |
| 72 | American Museum of Natural History | A | yes | yes | 40 | 8 | 6 | 0 |
| 73 | Trust For Public Land | C | yes | yes | 38 | 29 | 1 | 1 |
| 74 | Christian and Missionary Alliance | X | no | yes | . | 8 | . | 0 |
| 75 | Juvenile Diabetes Research Foundation Internati | G | yes | no | 87 | . | 23 | . |
| 76 | Christian Children's Fund | Q | yes | yes | 114 | 20 | 16 | 3 |
| 77 | Ducks Unlimited | C | yes | yes | 14 | 0 | 9 | 0 |
| 78 | Mental Health America | F | yes | yes | 12 | 13 | 0 | 2 |
| 79 | MakeAWish Foundation | E | yes | yes | 251 | 24 | 113 | 5 |
| 80 | Young Life | O | yes | yes | 5 | 4 | 0 | 0 |
| 81 | United Negro College Fund | B | no | no | . | . | . | . |
| 82 | Marine Toys For Tots Foundation | P | no | yes | . | 10 | . | 2 |
| 83 | Rotary Foundation of Rotary International | Q | yes | no | 66 | . | 3 | . |
| 84 | Catholic Medical Mission Board | Q | yes | yes | 4 | 11 | 1 | 4 |
| 85 | Trinity Christian Broadcasting Network | X | yes | no | 21 | . | 0 | . |
| 86 | Colonial Williamsburg Foundation | A | yes | yes | 67 | 25 | 10 | 4 |
| 87 | National September Memorial & Museum | A | yes | yes | 3 | 1 | 0 | 0 |
| 88 | The Conservation Fund | C | no | yes | . | 13 | . | 1 |
| 89 | Girls Incorporated | O | yes | no | 55 | . | 3 | . |
| 90 | WGBH Educational Foundation | A | no | no | . | . | . | . |
| 91 | Lincoln Center for the Performing Arts | A | yes | yes | 74 | 7 | 8 | 1 |
| 92 | Christian Aid Ministries | Q | no | no | . | . | . | . |
| 93 | United States Golf Association | N | no | yes | . | 3 | . | 1 |
| 94 | National Public Radio | A | no | no | . | . | . | . |
| 95 | Mercy Corps | Q | yes | yes | 56 | 27 | 7 | 1 |
| 96 | Project HOPE | Q | yes | yes | 7 | 4 | 0 | 1 |
| 97 | Museum of Fine Arts Boston | A | yes | yes | 0 | 21 | 0 | 0 |
| 98 | United Nations Foundation | Q | yes | yes | 85 | 29 | 3 | 8 |
| 99 | The Carter Center | Q | no | no | . | . | . | . |
| 100 | Kennedy Center for the Performing Arts | A | yes | yes | 168 | 69 | 9 | 15 |

NTEE categories: A (Arts, Culture, and Humanities); B (Education); C (Environment); D (Animal-Related); E (Health Care); F (Mental Health & Crisis Intervention); G (Diseases, Disorders & Medical Disciplines); H (Medical Research); J (Employment); K (Food, Agriculture & Nutrition); L (Housing & Shelter); M (Public Safety, Disaster Preparedness & Relief); N (Recreation & Sports); O (Youth Development); P (Human Services); Q (International, Foreign Affairs & National Security); T (Philanthropy, Voluntarism & Grantmaking Foundations); X (Religion-Related).

Appendix B.

Descriptive Statistics and Correlations

| | | Correlations | | | | | | | | | | | | | | | | | | |
|---|---|---|---|---|---|---|---|---|---|---|---|---|---|---|---|---|---|---|---|---|
| | Mean (S.D.) | 1 | 2 | 3 | 4 | 5 | 6 | 7 | 8 | 9 | 10 | 11 | 12 | 13 | 14 | 15 | 16 | 17 | 18 | 19 |
| 1. Facebook Presence | 0.65 (0.48) | 1 | | | | | | | | | | | | | | | | | | |
| 2. Twitter Presence | 0.73 (0.45) | .26 | 1 | | | | | | | | | | | | | | | | | |
| 3. Scl Media Presence | 1.38 (0.74) | .81 | .78 | 1 | | | | | | | | | | | | | | | | |
| 4. # FB Updates | 16.05 (15.09) | . | .27 | .27 | 1 | | | | | | | | | | | | | | | |
| 5. # Twitter Updates | 66.23 (65.74) | -.06 | . | -.06 | .64 | 1 | | | | | | | | | | | | | | |
| 6. Total # updates | 69.15 (73.04) | -.03 | .34 | .22 | .77 | .98 | 1 | | | | | | | | | | | | | |
| 7. # Dial. msg. (FB) | 2.52 (3.75) | . | .23 | .23 | .85 | .61 | .71 | 1 | | | | | | | | | | | | |
| 8. # Dial msg. (Twit.) | 8.6 (15.74) | -.05 | . | -.05 | .16 | .64 | .58 | .12 | 1 | | | | | | | | | | | |
| 9. Total # Dial msg. | 9.32 (15.65) | -.02 | .22 | .14 | .38 | .71 | .71 | .38 | .98 | 1 | | | | | | | | | | |
| 10. Fundraising ratio | 0.07 (0.11) | .06 | .13 | .12 | .06 | .06 | .10 | .14 | .06 | .11 | 1 | | | | | | | | | |
| 11. Lobbying Exp. | 43.20 (134.08) | .10 | .08 | .12 | -.02 | .29 | .26 | .09 | -.02 | .02 | .11 | 1 | | | | | | | | |
| 12. Prog. Svce. Rev. | 73.64 (272.08) | .04 | .08 | .07 | .11 | -.08 | -.03 | .07 | -.05 | -.02 | -.09 | .04 | 1 | | | | | | | |
| 13. ln(Size) | 18.87 (1.67) | -.06 | -.04 | -.06 | .14 | .04 | .08 | .03 | -.08 | -.05 | -.21 | .14 | .26 | 1 | | | | | | |
| 14. Website Age | 12.33 (2.06) | -.02 | .02 | .00 | .14 | .11 | .14 | .05 | .05 | .08 | .04 | .11 | .12 | .24 | 1 | | | | | |
| 15. Website Reach | 3.03 (0.54) | .11 | .20 | .19 | .30 | .40 | .41 | .19 | .17 | .22 | .11 | .33 | .13 | .19 | .29 | 1 | | | | |
| 16. Membership | 0.10 (0.3) | .11 | -.02 | .06 | -.07 | .04 | .01 | -.19 | .04 | -.01 | -.12 | -.04 | -.02 | .13 | .13 | .23 | 1 | | | |
| 17. Board Size | 34.28 (31.29) | -.06 | .03 | -.02 | -.14 | -.16 | -.13 | -.12 | -.10 | -.10 | .00 | -.01 | .01 | .22 | -.07 | -.08 | .02 | 1 | | |
| 18. Efficiency | 1.48 (6.21) | .07 | .06 | .08 | .11 | .07 | .09 | .15 | .04 | .08 | .84 | -.01 | -.02 | -.15 | .08 | -.03 | -.04 | -.01 | 1 | |
| 19. Donor Dependence | 0.48 (2.32) | .14 | .19 | .20 | -.07 | -.09 | -.06 | .02 | -.06 | -.04 | -.03 | .04 | -.03 | -.13 | -.01 | .23 | -.01 | -.19 | -.02 | 1 |
| 20. Gov't. Dependence | 0.14 (0.23) | .11 | -.05 | .04 | .12 | .20 | .16 | -.05 | .06 | .04 | .07 | -.08 | -.10 | .08 | .02 | .25 | .20 | -.02 | .11 | -.02 |